# A Discussion on the Chemical Origins of Life as We Don't Know It

Martin Rahm[a,*], Carol Cleland[c], Valentina Erastova[d], Robert Hodyss[e], Thanja Lamberts[f], Jonathan Lunine[e,g], Ishaan Madan[h], Dirk Schulze-Makuch[i], Rebecca J. Rapf[j], Paul B. Rimmer[k], Nita Sahai[l,m,n], Richard C. Remsing[b,*]

[a]Department of Chemistry and Chemical Engineering, Chalmers University of Technology, 41258, Gothenburg, Sweden

[b]Department of Chemistry and Chemical Biology, Rutgers University, Piscataway, NJ 08854, USA

[c]Department of Philosophy, University of Colorado at Boulder, Boulder, CO, USA

[d]School of Chemistry, University of Edinburgh, Joseph Black Building, David Brewster Road, Edinburgh, EH9 3FJ, UK

[e]Jet Propulsion Laboratory, California Institute of Technology, Pasadena, CA, USA

[f]Leiden Institute of Chemistry, Gorlaeus Laboratories, Leiden University, PO Box 9502, 2300 RA Leiden, The Netherlands; Leiden Observatory, Leiden University, PO Box 9513, 2300 RA Leiden, The Netherlands

[g]Division of Geological and Planetary Sciences, California Institute of Technology, Pasadena, CA 91125, USA

[h]Department of Earth, Atmospheric, and Planetary Sciences, Purdue University, West Lafayette, IN 47907, USA

[i]Institute for Physics and Astronomy, Technische Universität Berlin, 10623, Berlin, Germany; Department of Experimental Limnology, Leibniz-Institute of Freshwater Ecology and Inland Fisheries, 16775, Stechlin, Germany; Section Geomicrobiology, GFZ Helmholtz Centre for Geosciences, 14473, Potsdam, Germany

[j]Department of Chemistry, Trinity University, San Antonio, TX, 78212, USA

[k]Cavendish Laboratory, University of Cambridge, JJ Thomson Ave., Cambridge CB3 0HE, UK

[l]School of Polymer Science and Polymer Engineering, [m]Integrated Bioscience Program, [n]Department of Geosciences, The University of Akron, Akron, Ohio 44325, USA

**ABSTRACT:** The origin and nature of life beyond Earth remains one of the grandest open questions in science. While much research on this problem focuses on finding life as we know it – carbon-based, water-dependent, operating within a relatively narrow range of terrestrial conditions – far less attention has been paid to what life as we do *not* know it might look like. And yet this is one of the most fascinating and perhaps ultimately productive aspects of this core theme of astrobiology. This paper summarizes discussions held at a Telluride Science Workshop dedicated to the latter challenge: exploring the chemical origins of life unconstrained by terrestrial assumptions. Workshop participants, spanning chemistry, physics, planetary science, computation, and philosophy were asked to address seven guiding questions concerning the epistemology of speculative astrobiology, the future role of computation, the transportability of Earth-centric assumptions to alien environments, the prospects for low-temperature life (as on Titan), high-temperature life (as on Io), life under extreme pressure, and the conceptual bridge from prebiotic chemistry to astrobiology.

## INTRODUCTION

The question of whether life exists beyond Earth is among the most profound in science. Yet, our search for life has been largely guided by a single example – life on Earth and its underlying terrestrial biochemistry. This limited sample size of a single biochemistry raises the possibility that our definitions and detection strategies may be too narrow. A growing body of work in astrobiology asks whether life could arise through fundamentally different chemical principles in environments radically unlike those on Earth, leading to life as we don't know it.

We use the phrase "life as we don't know it" to refer to living systems that rely on a biochemistry that is of a fundamentally different chemistry than that of Earth life. This excludes Earth (or Earth-like) life that may exist, for example, at high salt concentrations or other extreme environments that nevertheless rely on the same fundamental Terran biochemistry. Instead, we focus on what has been called weird life – living systems that one may envision in the hot and acidic Venusian clouds, on the cryogenic and hydrocarbon-rich surface of Titan, or in other environments that are radically different to Earth. Studying the condensed phase chemistry and physics of these environments through the lens of astrobiology teaches us not only about the limits of life, but also how the essential molecular prerequisites for life might be realized through entirely different chemical mechanisms.

To address the potential for chemical origins of life as we don't know it, we convened a group of cross-disciplinary experts to push the boundaries of astrobiology and discuss thought-provoking questions over the course of a week-long Telluride Science Workshop in October 2025. What transpired was five days of completely open and creative discussions; no idea was dismissed as too weird, nor any environment too alien – the potential for scientific progress and advancing astrobiology led the way. In what follows, we present highlights of these discussions from a subset of the participants, which aim to identify specific directions, principles, and methodological developments needed for future investigations of condensed phase chemistry of potential relevance to life as we don't know it. Rather than digesting and rewriting the discussions as a single-voiced narrative, we preserve the identity of each contributor, although the comments are edited for clarity.

- **Martin Rahm and Richard C. Remsing**

## DISCUSSION

### Given the speculative nature of astrobiology, what qualifies as a scientifically good and valid question?

**Jonathan Lunine:** Astrobiology is no more speculative than that branch of cosmology dealing with the origin of the universe, whether there are multiple universes, etc., and so we should not be shy about posing daring questions. Nor should we be too ambitious and pose sisyphean questions …they must be answerable through carefully posed hypotheses experiment, observation that leads to modeling, experiment and observation that would pass the test for good science. Some of these tests may require decades and large amounts of resources, on the scale of major robotic missions to the moons of the outer planets or the return of samples from Mars. But in any such endeavor, the goal should be to do careful science at these destinations or on their samples, so that the questions posed are answerable[1]. Let us be bold in our curiosity, but feasible in our plans to satisfy it.

**Carol Cleland**: Our understanding of the nonliving universe is much greater than our understanding of life. More specifically, our experience with life is limited to a *single example* which (for all we know) may be unrepresentative of life considered generally. Biochemists have shown that familiar Earth life could have been *at least modestly different* in its basic biomolecules (proteins, nucleic acids), and lipids). We don't know how different a form of life arising in a very different chemical and physical environment (Titan?) might be. The question of the origin of life is ultimately a question about the environmental conditions necessary for life to emerge from nonliving molecules, and we are profoundly ignorant about how this transition occurred on Earth, let alone elsewhere. In contrast, our knowledge of stars, galaxies, and planetary systems isn't limited to a single galaxy, star, solar system, and planet. We have scientifically compelling stories, based on universal theories of physics (e.g., gravity and atomic theory) about how the first stars arose through gravitational collapse and nuclear fusion, burned out and gravitational collapsed, ... eventually forming third generation stars, composed of heavier elements, and solar systems with a diversity of different types of planets (including terrestrial planets, as well as gas giants). We have nothing similar in biology.

**Ishaan Madan:** A scientifically good question in astrobiology is one that practices disciplined imagination. One that is creative enough to expand what we think life could be but constrained enough that it can eventually be tested challenged or refined. A good question should lead to testable hypotheses that are falsifiable in principle, integrative in scope, method-generative, and evolutionary in structure. I agree with Jonathan that we must balance curiosity with feasibility. They need not be immediately answerable, but they should point toward observations, experiments, theory, or computation that could make them more answerable over time. Astrobiology sits at the intersection of countless disciplines, and its questions should reflect that conceptual expansiveness. The questions asked should compel us to revisit what we mean by life, intelligence, metabolism, information, replication, and perhaps even individuality. Valuable questions may also generate new methods, where they challenge us to develop better instruments, models, experiments, and frameworks for recognizing unfamiliar forms of organization. Consider a question like: can an alternative solvent sustain the concentration, templating, and catalytic functions that liquid water enables for Earth life? Such an exploration is falsifiable in principle, integrative across chemistry and planetary science, and immediately method generative. It demands new solubility frameworks, computational approaches to solvation, and experimental analyses. A good astrobiological question, simply put, is one that makes science (in any or all adjacent fields) more expansive.

**Carol Cleland**: It is very difficult to *scientifically* falsify hypotheses, however, (logically speaking) one can always salvage a hypothesis in the face of negative evidence by appealing to mistaken assumptions and poorly understood contingencies present at the time of observation/measurement. As an illustration, consider the failure of the Viking mission's GCMS to find any organic molecules on Mars, which was and (strangely) still is treated by NASA as falsifying the hypothesis that there is microbial life on Mars (despite the provocative but ambiguous results of the 3 Viking biology experiments); see Ref. [2] for a discussion. Subsequent missions to Mars detected organic molecules, contravening the evidence that supposedly falsified the hypothesis; the Viking GCMS wasn't (as erroneously assumed) sensitive enough to detect them. The point is the results of the GCMS didn't establish that the hypothesis 'there is microbial life on Mars' is false even though that is what it was said to have done. A similar situation will almost certainly occur if we encounter a form of life with a truly unfamiliar biochemistry. Ishaan conjectures that we can falsify claims about whether an alternative solvent, for instance, can sustain certain functions important for life on Earth. But just because conditions on Earth won't allow for something chemically doesn't mean that there aren't (poorly understood and unknown) conditions on other worlds that could. The truth is that the imaginations of theoretical chemists are Earthcentric. Even the most brilliant chemists don't have a good grasp on the possibilities for biochemistry in complex extraterrestrial environments unlike any found on Earth.

**Martin Rahm:** To help frame this discussion, I propose a "hot-crazy" scale for biosignatures, illustrated in Figure 1. By "hot" I mean a sign of life so clear that there can be little or no doubt; this is another life! Imagine, for example, the decisiveness of visual evidence of swimming creatures discovered beneath the ice crust of Europa or Enceladus, or the detection of living microbes 100 m below the Martian surface. Slightly less decisive, warm, could be the detection of skewed D/H ratios or enantiomeric excess in organic compounds found along the shores of Titan's hydrocarbon lakes.

The degree of "crazy", by contrast, reflects the effort required to obtain such evidence at a given time. In other words, how impractical, time-consuming, or costly it would be to test for the biosignature in question. In other words, the scale attempts to capture both the strength of the evidence and the difficulty of obtaining it. I wonder if such a scale would be practically useful. Let us play with the idea and maybe populate it with more concrete examples.

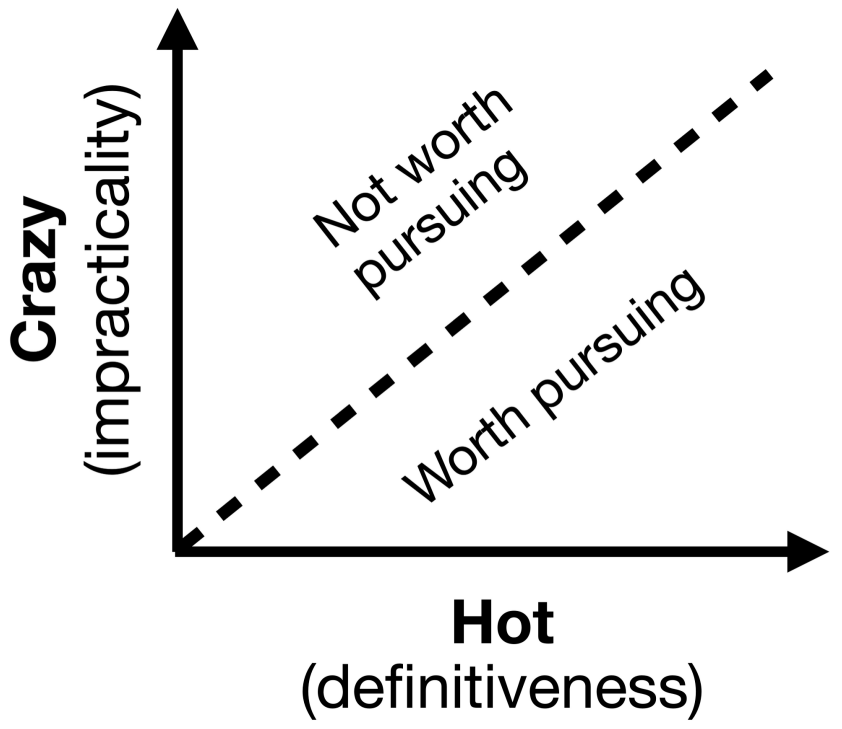

***Figure 1.*** *A hot vs. crazy scale for biosignatures.*

**Jonathan Lunine:** What do you define as time-consuming or impractical in general? To a theoretical chemist it might mean a model that requires too much time or computing power to run; to a team working on a multi-billion dollar mission to search for biosignatures on Enceladus it might mean something else, like, a mission concept that is impractical (a submarine in the ocean of Enceladus) or a flight time that is too long for many reasons including aging of the spacecraft and of the mission proponents.

**Ishaan Madan:** Building on Martin's and Jonathan's points, the hot vs. crazy scale is a property of biosignatures and of the observers evaluating them. What we consider "hot" or "crazy" depends on evolving thresholds for evidence, technological capability, risk tolerance, and imagination. In this sense, the scale is inherently dynamic. For example, a biosignature that is difficult to interpret or detect today may become more compelling as observational tools and planetary context improve. The scale might therefore benefit from a third dimension of fruitfulness; a dimension that captures the broader scientific knowledge gained from a search even if it does not yield a direct detection of life. I hope to see such a scale that is less about strict ranking and more of a way to calibrate ambition, evidence, feasibility, and scientific return on investment.

**Martin Rahm:** Yes, the meaning of "crazy" (the y-axis) will inevitably change over time. It represents the balance between expected payoff and the resources required, weighted relative to other avenues. On definitions, I can tell you what a "hot" biosignature need not be: it need not involve molecular complexity.

Molecular complexity is something inevitably brought up when discussing biosignatures (see for example, Sharma et al[3] and the ensuing debate[4], or Neveu et al[5], Barge et al[6]). It is probably uncontroversial that a sufficiently complex molecule - quantified in some way - that is found in abundance would be a "hot" biosignature. The difficulty is that detecting and characterizing complex molecules or materials far away is often "crazily" challenging: too technologically demanding or expensive. A case in point is Venus. None of the many currently planned missions to Venus are well suited to detect polymers or other complex macromolecules, even though such species could be present in the planet's potentially habitable cloud layers.[7]

I will take this opportunity to argue for something different: structurally simple molecules can also be very "hot" biosignatures.

Consider, for example, the hypothetical tetrahedral $N_4$. This is a molecule that good quantum chemistry predicts to be very kinetically stable (persistent even at high temperature) but also so energetic that ~13.5 MJ/kg (760 kJ/mol) is released upon its transformation into $N_2$.[8] Imagine if we went to Titan and discovered a pool of concentrated $N_4$. Now that would be extraordinary. I spent several years attempting to make $N_4$ with Karl Christe using elaborate low temperature glow-discharge experiments. That (unpublished) experience was sufficiently humbling that I can assure anyone reading that the detection of concentrated $N_4$ anywhere beyond Earth would constitute a very "hot" biosignature.

The argument is based on thermodynamics: a molecule that rich in energy cannot plausibly accumulate in bulk through abiotic processes; it simply lies too drastically away from equilibrium with its surroundings. Of course, out-of-equilibrium chemistry as biosignatures is not a new idea.[9] However, many classical examples, such as $O_2$ – if it was discovered around an exoplanet – are ambiguous because one cannot rule out abiogenic explanations. The detection of something like $N_4$ would be different. In fact, given the difficulties involved in its manufacture under controlled conditions, I would argue it is a potential chemical technosignature: a simple molecule whose production requires advanced technology.

Admittedly, one cannot rule out the possibility that such compounds might be produceable by some form of simple cryogenic life in need of explosives; a scenario I will address when answering a later question. Either way, $N_4$ and other energetic molecules exemplify a class of potential biosignatures that are decidedly not complex.

More broadly, one can imagine many reasons why life might produce structurally simple molecules or materials, that nonetheless are "hot" biosignatures. At low temperature it might, for example, be challenging to create multiple bonds, or even move larger molecules around. It could then be advantageous to synthesize simpler molecules with innate complex electronic structures instead, e.g., with the ability to persist or evolve in excited electronic states. Finding monodispersity in structurally simple polymers is another example of a "hot" chemical bio- or technosignature.

**Dirk Schulze-Makuch:** I find this idea intriguing, and I think we need to distinguish between biosignatures and technosignatures. In general, you need complex molecules to do complex functions, as pointed out by the late Bob Shapiro[10]. To put it starkly: you can't have life just with CO2, $CH_4$, or similar small molecules. While something like N4 might have been produced by life as a fuel source, creating and stabilizing such a molecule is complex, and therefore would require life composed of complex molecules. Therefore, if concentrated $N_4$ is found, are there still large macromolecules in addition to $N_4$? What I would like to see in this case is an enrichment in specific high AMU numbers[11], which relates to specific macromolecules doing work (but these high numbers would not correlate necessarily to Earth's life distribution, where it presents certain amino acids, nucleotides etc).

**Nita Sahai:** Following on Martin's and Dirk's comments, Teflon and similar complex molecules could then be a technosignature.

**Paul Rimmer**: What features of the energy landscape would be convincing enough to demonstrate that one of these energetic molecules is a good signature? Is it a fuzzy boundary?

**Carol Cleland:** The problem is that we don't have a universal theory of life for inferring which characteristics of familiar life are common to all life (wherever it is found in the universe) and which characteristics are abiotic in the sense that living things never have them (regardless of the environment in which they are found). Characteristics had by living things on Earth may be maladaptive in certain extraterrestrial environments, and vice versa; the characteristics of living things are coupled to their environments through adaptation, and hence one would expect that life found in strange environments would have very different characteristics from familiar life. This makes coming up with a set of reliable "biosignatures" for searching for unfamiliar forms of life very problematic; for a characteristic to be a biosignature it must be indicative (but not necessarily defining) of life.

In this context, it is important to appreciate that I am not saying that it isn't valuable to formulate "hot" and "crazy" biosignatures. What matters is how they are deployed in a search strategy for life as we don't know it. The strategy that I recommend is nontraditional: Search for potentially biological anomalies (vs. life per se) using tentative (vs. Defining) criteria. Tentative criteria would be based on what we know about both living and nonliving things on Earth, plus theoretical extrapolations of the sort discussed in this paper. The important point is not to treat tentative criteria as definitive for life and to use characteristics thought to be exclusive to abiotic systems as well. The basic idea is to design a search strategy for a *particular environment* using a suite of tentative criteria suggestive of what life and nonlife might be like i*n such an environment*. If one is lucky, these criteria will be satisfied, and life will be discovered. If the environment is significantly strange, however, one is likely to encounter a weird form of life exhibiting ambiguous characteristics (from an Earthcentric perspective) suggestive of life but also of nonlife. Such a phenomenon would defy classification as living or nonliving. Traditional search strategies would thus reject it as a candidate for life. Yet an ambiguous (aka anomalous) phenomenon is just what one would expect of a truly strange form of life adapted to an environment different from Earth's. For more detail on how one could apply this proposal, see Cleland and Wong,[12] Cleland and Rimmer,[13] and Cleland (Ch. 8).[14]

**Dirk Schulze-Makuch:** Similarly, finding large quantities of chlorophyll and/or similar molecules would be a signature of photosynthetic life. Also, something brought up by Carol: one of the main problems is that it is not well defined or known what abiotic processes could produce false biosignatures.

**Valentina Erastova:** Carol's anomaly-driven approach has a practical consequence for mission design: the site optimized for finding organics in abundance is not necessarily the site optimized for finding biological anomalies. Maximizing the chance of detecting any organics would be going to the richest organic environment available. But if the goal is to identify something that *should not be there* (e.g., a peptide with non-random sequence, or a polymer with anomalous chain-length distribution, or a molecular abundance unexplained by the modelled abiotic production/loss pathways) then one should look where the background abiotic chemistry is best characterized and most predictable. These are different target criteria and conflating them risks optimizing a mission for the wrong detection problem.

**Thanja Lamberts:** Given the speculative nature of the field, it might be worthwhile to look across the borders of disciplines and borrow from other fields that are speculative to draw from how they – in early days – have found ways to construct good and scientifically valid questions. This may be fields closer to home, like cosmology or exoplanetary research, or even further from it like behavioral economics or historical linguistics.

**What role would you like computational astrobiology to play in the decade ahead?**

**Martin Rahm:** To help frame this question, it would help to first attempt a definition of what *computational astrobiology* is. In the broadest sense, I think that "the testing of astrobiological hypotheses using computational tools" does the job, would you agree? That definition would encompass work done with quantum chemistry, molecular simulations, and machine learning, etc., if those simulations are aimed at answering life-beyond-earth questions. One could then debate whether more conventional modeling of prebiotic chemistry (life as we know it) fits that definition or not. Either way, I think computational astrobiology is a field of research likely to grow in importance with the advent of more advanced, more reliable computational methods, but one that I foresee will, at times, struggle with unfounded speculation and poorly constrained questions. A non-exhaustive list of criteria for "good practice" might be:

Maintain focus on well-defined and limited questions that can be answered by computation to some degree of certainty. For example, could this hypothetically astrobiologically-relevant material form in a given extraterrestrial environment? Could it perform a particular kind of catalysis, absorb light at which frequency, float on a given liquid, or facilitate that kind of hypothetical metabolic cycle?

Disclose method limitations honestly. Given the interdisciplinary nature of astrobiology, a computational chemist or physicist needs to be especially clear in communicating the caveats and limitations of any predictions[15, 16]. These will not be obvious to expert planetary scientists or geologists unfamiliar with the computational methods in play.

Provide clear distinctions between computational results and speculation. Jonathan is right that astrobiology is no more speculative than some branches of cosmology, but I would still maintain that our field is, on balance, more speculative than most other scientific disciplines. Speculation comes with the territory and is often necessary. However, too much unfounded speculation can be harmful, to the clarity of communication, to the public perception of scientific progress, and ultimately to the advancement of the field itself.

Provided such good practices are upheld more than they are not, the field of computational astrobiology could mature into something with real impact on how the limits of life are explored, and life detection.

**Valentina Erastova:** A recurring challenge in computational astrobiology, and equally in experimental prebiotic chemistry, is the mismatch between accessible and geochemically relevant timescales. Computational and laboratory experiments are both models of a chemical environment: one samples chemical space through simulation, the other through controlled perturbation in the laboratory. Both are constrained by detection limits and by the timescales over which reactions/processes can be observed. A reaction that is too slow to yield detectable product in a laboratory setting, or a process too slow to converge in a simulation, may nonetheless contribute to prebiotic chemistry if products accumulate faster than they degrade or are removed, within the lifetime of the relevant environment. These technical limitations can bias our assessment of geochemical plausibility, if we privilege processes accessible to direct simulation or laboratory measurement. Although kinetic models can propagate reaction networks over geological timescales, their predictions remain conditional on network completeness, rate uncertainties and the assumed environmental history.

Consider mineral-mediated peptide formation: what matters is not whether polymers form, but whether they form with sufficient selectivity over sufficiently long periods – conditions that are exceedingly difficult to replicate or even approximate in the wet lab. Equally, understanding how these polymers decompose or transform over geological time is as informative as understanding their formation.

**Martin Rahm:** Here I must partly disagree with Valentina. Besides observation, computation can be helpful in evaluating chemical outcomes and basic feasibility over very different timescales. Time-resolved quantum chemistry (molecular dynamics) can at present be run for only a few nanoseconds, while metadynamics and similar approaches allow the sampling of processes occurring over seconds or minutes. For well-defined elementary steps - including ones never observed - individual barrier heights and energies can often be computed at quite high accuracy, then knit together into reaction networks whose kinetics can be propagated to arbitrary timescales. This decouples the timescale one can sample from the timescale one can reach: we never simulate a century, but solve coupled rate equations out to one. We recently showed, for example, how suggested mechanisms for forming the nucleobase adenine from hydrogen cyanide likely switch several times in proceeding from minutes to months to centuries.[17] Such kinetics modelling is never certain but can also draw on a combination of experimental and computed reaction rates. In fact, doing so is fairly standard for the (approximate) modelling of astrochemistry and the evolution of entire solar systems.[18]

**Dirk Schulze-Makuch**: I´m not as enthused about computational astrobiology. Yes, it can be helpful, but what you get out is mostly determined on what you put in. Do you then really expect creative solutions? I think our brains are at this point still much better in that.

**Rick Remsing:** Our brains and theoretical/computational astrobiology are not separate entities working independently. Proper application to problems in astrobiology rely on building models based on the relevant physics and using theoretical and computational techniques to predict the interesting behavior that emerges from the physics. In that sense, yes, the possibilities you get out are determined by what is put into the model, in the same way that what can be measured experimentally is determined by what is put into the experimental setup (reagents, temperature, pressure, etc.). As a result, theory and computation offer the opportunity of creativity and understanding how emergent behavior is dictated by the fundamental physics and constraints imposed on a system. Of course, real systems are often more complex than those studied computationally, creating more opportunities for surprises like unexpected reaction products, but in principle, the same surprises are present in computational models if a similar level of complexity is included.

Theory and computation-driven research created unexpected insights into astrobiology, such as identification of overlooked reaction pathways[19-22] and unexpected aggregation and reactivity in prebiotic solutions[23, 24], as well as the development of theoretical frameworks like assembly theory[3, 25] and fluctuation theorems that constrain the behavior of living systems[26-28]. Additionally, extremely cold or hot environments on the surface of many planetary models are often prohibitively difficult to mimic in laboratory settings, making computation the only viable probe of chemistry at these extremes.

A significant risk to understanding life as we don't know it is relying too heavily on intuition drawn from terrestrial biochemistry and extrapolating to regimes beyond its applicability. Instead, building physics-based models enables one to think about the chemistry of life as we don't know it in a way while ideally lessening bias.

**Nita Sahai:** Based on my own research experience with computational chemistry approaches (ab initio cluster and classical force field MD simulations) to understand mineral-organic systems in biomineralization and prebiotic clay-catalyzed RNA polymerization, the possibility of getting inaccurate results from computational studies is very high. Hence, benchmarking to experimental results and ensuring accuracy of force fields, including inclusion of mineral-organic or mineral-water interface cross-term parameters, etc., are critically necessary for accurate predictions.

**Jonathan Lunine:** I will answer this for glamorous computational astrobiology (GCA), as distinct from (e.g.) finding rate constants for reactions which might be called utilitarian computational astrobiology or UCA. GCA should provide answers to questions that in principle are addressable by experiments but for which the range of parameter space needed to be covered is too large (and/or the number of adjustable parameters too high) to be fully covered by experiment. GCA is the tent that is anchored securely to the ground by the stakes of experimental work. It should allow the intellect to soar in exploring in a physically and chemically rigorous way different possibilities for molecular systems that are separated from us in physical conditions (Titan surface), in time (ancient Martian ecosystems) or in ontological sequences (steps in the transition or transitions from prebiotic systems to biology).

I add that the 2030's will hopefully be the decade of the large moons of the outer solar system, with three major missions collecting data in the early-to-mid-2030's:

Europa Clipper: in flight to Europa; arrives 2030[29]

JUICE: In flight to Ganymede and Callisto; arrives 2031[30]

Dragonfly: Undergoing final preparations; will be launched to Titan in July 2028 for arrival in 2034.[31]

GCA and UCA combined with experiments will be needed to interpret results. Five to nine years may seem a long time, but it is really just two or three cycles of grant proposals. We haven't talked much about Ganymede and Callisto, and they may well not hold habitable environments, but if they have deep oceans then the high-pressure life question comes into play. Europa, of course, might support Earth-like life, but we will hopefully know more about its near-surface ocean in five years or so.

GCA has demonstrated its utility in modeling possible chemical evolution at the surface of Titan, which is ideal because of the exotic nature of the conditions coupled with our knowledge of the surface conditions and atmospheric composition.

If the azotosome (a hypotetical cell-membrane analog build from nitrogen-bearing organic molecules rather than lipids, proposed for the hydrocarbon seas of Titan) is energetically disfavored, what would spontaneously form the equivalent of vesicles in a Titan sea[32]?

Are there coupled polyacetylene-polyimine structures that could be of interest[33]?

Given that reactions seem to be facile based on recent studies of tunneling at Titan temperatures[34] what chemistry might happen in the seas or on the surface? There is already a disconnect between the predicted abundances on the surface of species made by chemistry in the atmosphere, and the actual detected abundances. Might chemistry be altering the organic composition on the surface?

**Rob Hodyss:** While computation and theory are often used to do exploratory work and make predictions to be tested, these techniques can also be used to compute quantities that can be prohibitively difficult to measure experimentally, such as gas phase reaction rates and solubilities in cryogenic solvents. Quantitative predictions of these quantities can be fed into large-scale models, enabling results that would be impossible to obtain otherwise.

**Ishaan Madan**: I agree with Rob and Valentina. Computational astrobiology can be a valuable first-pass filter for deciding where experimental time, mission resources, and theoretical attention should be directed. In many, arguably most, extraterrestrial environments, the relevant chemical inventories are poorly constrained, the geological context is not understood globally, reaction networks are incomplete, rate constants are unavailable for thousands of possible chemical species and reactions, and direct experimentation is difficult or even impossible. Computation can help map what is physically or chemically permitted before we ask what is kinetically realized. This is important because elucidating even one chemical pathway can be extremely resource intensive. Experimental studies often focus on one pathway, or a limited subset of pathways, and usually cannot operate on geologic timescales. Computational work faces a similar challenge where mechanistic studies can require years of work to understand even one reaction network or a limited chemical system. Scaling these to thousands of potential species in a complex soup of organics in varying environments is unrealistic. Therefore, computational astrobiology can help narrow the possibility space by identifying which chemistries, environments, and questions are worth pursuing further.

There is also an educational component. As computational methods become more central to astrobiology, we need to teach students how to run the calculations and, more critically, how to think alongside them. That means understanding the assumptions, limitations, uncertainty, and the difference between a model result and a claim about nature. Done well, computational astrobiology has the potential to train the next generation to be simultaneously imaginative and disciplined, willing to explore radical chemical possibilities while remaining grounded in the laws of physics, chemistry and testable prediction.

**Valentina Erastova:** To assess the planetary relevance of computationally proposed chemistries, we need targeted experiments that examine their formation, stability and reactivity under relevant conditions. Carbon-free heteropolymers proposed for Venus's cloud layers provide a target for such testing.[7] The azotosome hypothesis illustrates how subsequent calculations and experiments can constrain predictions of membrane stability and formation.[32, 35, 36] Closing this gap requires both greater methodological reliability in computation (so that predictions carry sufficient confidence to guide experiment) and a stronger culture of experimental replication across groups. Neither alone is sufficient. More reliable calculations reduce the cost of prioritizing which experiments to run, while independent experimental replication provides the ground that validates or redirects the computational efforts.

**Thanja Lamberts:** I would like to advocate first and foremost for a cleaner use of the categories within computational astrobiology, in a way similar to how things are perceived in astrochemistry. Molecular-level modeling, often aided by the aforementioned force fields, density functional theory, or other methods with a strong physical-chemical backbone is one thing. I agree with Ishaan here that it is not trivial to study multiple processes accurately, especially not when taking benchmarking into account. Intermediate scale modeling, where some physical chemical processes are taken into account and the models run over large timescales is another thing. Often these types of models are necessary to bridge the short chemical timescales to the long timescales of relevance in interstellar/astrobiological contexts. And finally, scaling up to more physics-heavy models where, e.g., impacts are explicitly taken into account is yet another branch.
Each of these three offers a different range of creative possibilities to play with, in some cases it might be possible to change pressure, temperature and starting chemical which has the benefit that it offers the possibility to explore, and the disadvantage that the choice of underlying assumption on, e.g., rate constants may be crucial. The advantage of the molecular-level models is we do expect life as we don't know it to follow the same physics and quantum mechanics as life as we do know it, even if we cannot study it experimentally.

**What properties presumed necessary for Earth-life may not apply for life as we do not know it?**

**Jonathan Lunine:** Hydrophobic or hydrohilic structures; strong polar interactions; presence of the element O in compounds, or other chemical properties necessary for molecules to function in the presence of liquid water.

**Ishaan Madan:** Many properties we assume to be necessary for life may very well simply be properties of life as we know it and may not be retained beyond Earth. Life beyond Earth may not require water specifically; rather some medium, interface, or phase structure that allows matter, energy, and information to be organized and transformed. The same may be true of carbon. Carbon's versatility makes it uniquely suited for Earth biochemistry, but life elsewhere may exploit different chemical logics in environments where carbon chemistry is not the most optimal route.

Our assumptions about replication and information storage may also be too narrow. Earth life stores information digitally through DNA and RNA, but synthetic biology already shows that heredity and evolution may not be exclusive to those two polymers. Xeno-nucleic acids (XNAs) have been shown to store and recover genetic information, and some XNA systems can undergo Darwinian evolution and fold into target-binding structures.[37] If generalized, this suggests that heredity and evolution may not be features unique to DNA or RNA. Information storage in life elsewhere may also transcend the specific hydrogen-bonding scaffolding used by terrestrial nucleic acids.

Other assumptions about metabolism and disequilibrium may not be held either. We are used to life with the ideas of fast, continuous, and strongly out of equilibrium metabolism. However, life elsewhere, any other life, may operate on timescales radically misaligned with ours (e.g., slow life on Titan, as discussed later). Similarly, life elsewhere may not need to avoid equilibrium in the same way Earth life does. One could imagine life elsewhere moving between dormant and active states, maintaining local pockets of order, or exploiting rare environmental fluctuations rather than relying on continuous metabolic output. I see this analogous to hibernation on Earth, to a limited extent since metabolic activity is reduced rather than halted, but it does gesture toward a broader possibility.

I also think we should distinguish the conditions needed for the emergence of life (urability) from those needed for the persistence of life (habitability), since these are related but not identical terms.[38] Origins of life may require localized, dynamic, and chemically productive settings (e.g., gradients, concentration mechanisms, wet–dry/freeze–thaw cycling, mineral interfaces, etc.). A system may need fluctuations to become alive but then need more stable conditions to remain alive and evolve.

Finally, life elsewhere may not leave familiar biosignatures, and its (coded) purpose may not be just survival and reproduction.

**Martin Rahm:** If one allows for the possibility of hypothetical solid-state or electronic life, one realizes that most criteria often taken as essential for life are optional. There is then no need for liquids, membranes, polymers. Even evolution might not exist in the conventional sense, but instead be replaced by individual adaptation, change-of-self. Solid state life might also be distributed over many locations, or exist as disjoint units that can be active or inactive. Supposed requirements of life needing spatial constraints (as in membranes or bodies) could then go out the window or become blurred, depending on how one defines things. What properties are then common to all life? Probably functional information storage and processing, the ability to do work, replication, growth, or at minimum the ability to do self-maintenance.

**Dirk Schulze-Makuch:** Perhaps it is easier to ask what has to be retained in order for us to call it life:

I think it must be in thermodynamic disequilibrium with its environment (otherwise its dead), which is on Earth accomplished with membranes but of course it could be something else.

It has to have some way of reproduction otherwise it will be extinct very soon. That does not need to be DNA or even be chemical in nature.

It has to be able to do some work and react or adapt to the environment as outside environments are always changing. Work is done on Earth by proteins. Doesn´t have to be proteins but something has to do something.

This opens up a huge variety of possibilities. For example, life could use different types of energy sources other than exploited on Earth, such as magnetic or kinetic energy, or thermal energy, or even osmotic or pressure gradients. It could use different building blocks like silicon in a radically different environment and also different solvents such as sulfuric acid or methanol[39].

**Nita Sahai:** Do we need some boundary, maybe one created by phase separation, such as liquid-liquid phase separation[40]?

**Rob Hodyss:** Maybe the boundary doesn't need to be a chemical one, but can be purely physical: a puddle, or a cavity in a rock.

**Ishaan Madan:** There needs to be a mechanism for concentrating reactants.

**Rebecca Rapf:** The presence of a boundary or an interface seems to me to be quite important. Such interfaces not only provide separation from the surroundings (i.e. a unit of life), but the boundary between two phases (e.g. air-water, lipid-water, solid-liquid, liquid-liquid) also provides an interesting reaction environment that differs from bulk phases.

Such interfaces can also provide a driving force for the self-organization or self-assembly of molecules. For instance, air-water interfaces align and concentrate organic molecules as well as altering the thermodynamics and kinetics of reactions compared to bulk aqueous solutions [41]. Observations of changes in reactivity in aqueous microdroplets are currently an area of intense research activity[42-44], including the potential relevance of these reaction environments to prebiotic chemistry for life as we know it. Interfaces also facilitate gradients (e.g. concentration, pH, redox, etc), which are also important when we consider driving forces for prebiotic chemistry leading to either life as we do or don't know it. While I think we can take a broad view of what such interfaces might look like, I agree with Nita that a boundary of some type is likely to be necessary/universal.

**Nita Sahai:** On a different note but one that is related to both the question of what constitutes a good question in the Astrobiology field as well as to the current question being discussed in this section, I would like to bring up a more philosophical concept here. That is, to address the importance of ontology versus ontics in Astrobiology and Origins of Life studies.

The age-old difficulty in providing an Ontological account of Life has led to a philosophical retreat in the Astrobiology field to adopting the Ontics approach of defining the behaviors of life. The biochemical ontics approach defines life as entities capable of reproduction, self-replication, metabolism, etc.[45], while the "principles of life" or physical chemistry ontics description includes disequilibrium, self-organization fueled by a constant energy flux, complexity, emergence, etc.[46-53] However, various life-life behaviors are seen in systems that we would not generally recognize as "living."

For example, the idea of chemically complex molecules as potential biosignatures quantified, say, by the minimum number of steps needed to synthesize them[3], was shown recently not to be unique to organic molecules and may apply to some minerals also[54]. Minerals (naturally occurring inorganic crystalline compounds with fixed chemical composition or composition varying in a solid-solution) also possess properties of being able to "self-replicate" and "grow," with "memory" in their crystal structures, defects as "variation" across generations, and to "sense and respond" recursively to changes in the external environment, as reflected in the crystal's chemical composition changes. These are not new ideas, rather, they harken right back to the very early studies of the origins of life[55-59]. Minerals as life is an unpopular idea today but, if we are to consider Martin's ideas of solid-state biochemistry in Life as We Don't Know It, especially as in very cold environments, then the distinction between Life-as-We-Don't Know-It and Life-as-Minerals becomes blurry just as the distinction between Life-As-We-Know-it and prebiotic organic molecular systems is blurry. Furthermore, polyoxoanions also have the ability to self-replicate, grow, sense their environment, exhibit chiral selectivity for other inorganic ions, and even for organics such as amino acids[60-62]. A final example points to synthetic organic crystals and metal-organic frameworks (MOFs), that are capable of sensing and responding to environmental stimuli, retaining shape-memory, etc.[63-65] Together, these inorganic, organic-inorganic and purely organic systems, that show partial life-like behaviors, can also be persistent in the sense of being kinetically-trapped as long as energy flow is maintained regardless of their specific chemistry. So, if our goal is to understand Life and its Origins for the purposes of astrobiological detection, regardless of whether it is as-we-know-it or as-we-do-not-know it, then these alternative systems mentioned above could arguably also be considered as life.
In my comments here, my aim is not to argue for or against any of these particular systems as being life. Rather, I am attempting to draw attention to a philosophical aspect of the scientific investigations of Life and its Origins. I would submit that both the biochemical ontics approach and the physical chemistry ontics approach bring us back to the need for an Ontological understanding of Life.

I realize that this is likely the minority-opinion in our workshop, but I submit that science and philosophy are inseparable aspects of understanding Life, its Origins, and its Detection. In this regard, the conception of life (pun intended) as a Kantian whole[53] is at least one critical aspect to include in an ontological description of Life, and a dual-aspect monism or objective analytic idealism would provide the philosophical basis for a complete ontology[66, 67].

**Jonathan Lunine** In this regard, the further exploration of different environments on Titan is crucially important depending on what molecules and structures formed by such molecules are found there.[68]

**Rob Hodyss:** Nita's point about the complexity and properties of minerals and other crystalline materials is particularly relevant for Titan. Our recent work[69] has shown that small organic molecules on Titan's surface can form a variety of co-crystals, and we expect that there are many more yet to be discovered.

**Nita Sahai:** Where/how do you differentiate life from crystal chemistry?

**Rob Hodyss:** It's good that it's hard to distinguish between life and crystal chemistry, for example, because this means that you can transition from one to the other.

**Paul Rimmer:** Life as we know it is modular. It is built from interchangeable parts or building blocks. For any molecular building block, there are a set of other molecules with which it could be exchanged. The probability of any building block being located in a particular part of the macromolecular structure must be chemically similar to the probability of any other molecule with which it could be exchanged. Life as we know it is cellular, and each cell has macromolecules, that generate boundaries between inside and outside the cell (membranes), control the chemical kinetics inside the cell (amino acids), and store and copy the information that expresses the properties of the cell (DNA and RNA). Each of these is made of specific building blocks. Lipids form membranes. Amino acids form proteins. Nucleotides form RNA and DNA. Metabolites are passed from one pathway to another, converting environmentally available molecules into these building blocks, and storing light or other forms of energy, e.g. chemical energy, often in the form of ATP and GTP – the same modules used for DNA and RNA.

Reactions between these molecules are thermodynamically disfavored in water. They are dehydration reactions; see Figure X. In the right local context, they react. Often this local context is set by a specific protein, which will change the local water activity to allow the desired reaction to proceed. Much of prebiotic chemistry has been focused on finding physical or chemical conditions, that allow dehydration reactions to proceed within a natural environment. Candidate physical conditions include wet-dry and freeze-thaw cycles. Candidate chemical conditions center around condensation reactions, driven by molecules like hydrogen cyanide, cyanoacetylene, and carbonyl sulfide. See [70] for a useful review of several key biochemical dehydration reactions and proposed prebiotic alternatives.

Life as we do not know it could differ from Earth life in several ways. It could use different chemical building blocks, so that the modules are not amino acids, nucleotides, sugars, and lipids. It could use different chemical mechanisms, so that catalysis, information copying, and energy transduction do not resemble enzymes, nucleic acids, and proton gradients. It could operate in different temperature and pressure regimes. It could use different solvents, and alternative solvents should be judged by their own physical and chemical logic rather than by how closely they imitate water [71].

It is imaginable that life could even use different elements. Arsenic in place of phosphorus [72] or silicon in place of carbon [73]. But these substitutions face real chemical limits, and in water at least, are unchemical (the chemist's equivalent of unphysical). Arsenic and silicon do not simply do what phosphorus and carbon do in water. But it could be imagined that arsenic and silicon, or other elements, could do in a non-water solvent what phosphorus and carbon do in water. The limits of chemical possibility may be less absolute for these elements if they found themselves in a different solvent.

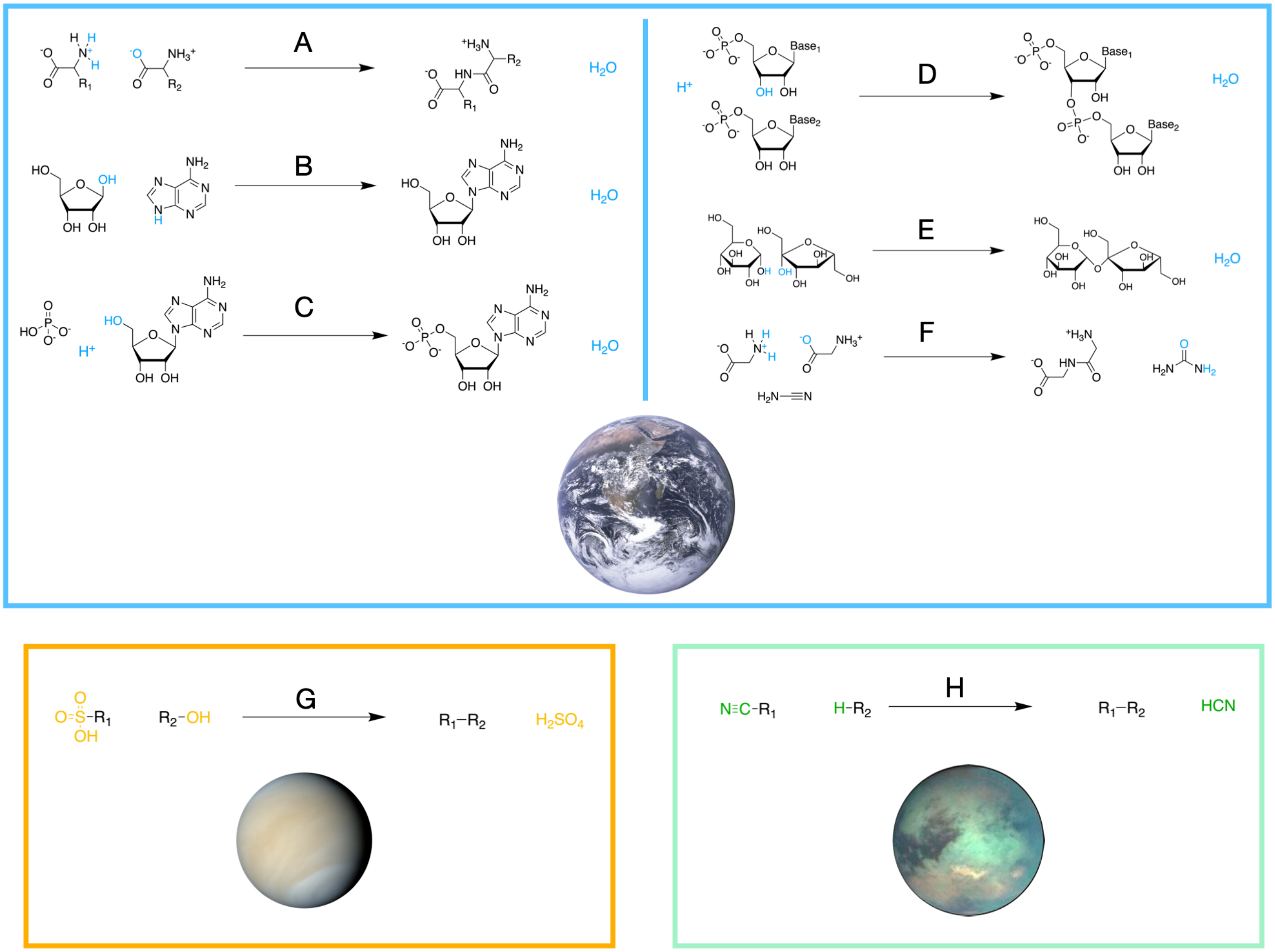

***Figure 2.*** *Reactions with life as we know it are dehydration reactions in water, providing chemical control. And then there are two hypothetical reactions, one for clouds of Venus, and one for liquid HCN, which would be broadly analogous to the reactions we expect in water; the solvent affords chemical control, if the building blocks have the right chemistry for it.*

As mentioned before, every single step toward building biochemistry from abiotic sources involves dehydration: peptide bonds, phosphodiester bonds, glycosidic bonds, and many ester linkages involve expelling water from the molecular constituents into water. This water problem for prebiotic chemists [74] may also be a clue. Earth life uses modules whose spontaneous reaction with one another is disfavored within the solvent where they swim.

That point matters for alien life. First, it is difficult to imagine life of any chemical constitution without modularity. What would be copied, and how? And if nothing is copied, there is no self-replication, and what would life be without self-replication? I leave these as open questions for the reader, and choose instead, at this point, to make a choice of focus, on life (as we know it, as we don't) that is modular. Next, we might ask, could modular life persist if the modules freely reacted with each other? I think not. If a solvent makes every possible module react with every other possible module, then chemical control is lost, and any kind of stable emergent behavior, such as kinetic dynamic stability [75] will be impossible [76]. The system becomes a tar [77], a clutter [78], or a short-lived reaction mixture [76]. Chemical control seems cleanest when the modules of life, any life, find mutual reactions thermochemically uphill when the activity of their own solvent is high.

This is also where chemical kinetics becomes the bridge between chemistry  in the lab and chemistry in the wild [76]. It is not enough to ask whether a reaction is possible. One must ask whether it occurs fast enough, selectively enough, and in the right sequence under the environmental conditions being proposed. In the carboxysulfitic scenario, a relevant candidate prebiotic scenario for life as we do know it [79], rate constants and rate laws change with pH, light, and chemical context; the same nominal ingredients do not imply the same network behavior everywhere [80]. That is the lesson I would carry into life as we do not know it. A proposed alternative biochemistry should not be evaluated only by listing possible building blocks. It should be evaluated by asking which reactions dominate, which reactions are suppressed, which intermediates persist, and which environmental features provide the potential for the system itself to grab control of its own chemical steps.

Prebiotic chemistry, whether leading to life as we do or do not know it, is "do-nothing" chemistry [76]. A natural environment does not get to purify intermediates after each step unless the environment provides a straightforward path to purification. It does not get to add reagents in a preferred order unless the environment itself favors that order. It does not get to remove side products because a chemist knows they are inconvenient. In other words, the environment selects for structure over function. Multistep chemistry therefore faces the arithmetic demon: modest losses at each step become catastrophic over many steps, and side reactions quickly erase the desired pathway unless some ordering principle is present [76]. In Earth-like prebiotic chemistry, that ordering principle might be light, pH, drying, freezing, surfaces, concentration, or selective degradation. In more alien chemistry, it might include something else, like quantum tunnelling[34, 81]. But the ordering itself must be there.

Solid-state life poses a riddle for this proposed rule of selective modularity. Life seems to need some medium that allows parts to move, sample alternatives, and be rearranged in a way that, eventually, is selected for its function, rather than its structure. Solvents can do this very naturally. Solvents let modules encounter one another, separate again, and react only when the right conditions are present. A solid can preserve structure well, including modular structure, but it may preserve that structure too well. It can make chemistry depend too strongly on nearest neighbors, lattice positions, defects, and long-persistent elemental substitutions. How would these conditions allow selection for function? In an informational polymer, for example, it would be fatal if one letter too strongly dictated the next: if A strongly preferred G over U, the sequence would already be biased before function could select among alternatives [82]. The same problem becomes sharper if nearest-neighbor structure determines most of the chemistry. Then the system may not be free enough to explore functional space. It may only explore the space allowed by packing. A solid-state life would therefore need some way around this. Perhaps selection for function could occur through defects, surfaces, pores, transient melts, boundary conditions, or mobile molecules trapped in solid latices. If so, this would take place on a vastly different timescale than life as we know it. Maybe quantum mechanics can help out here? Probably a workable model of solid life, if such a thing exists at all, would be much stranger than anything I can presently imagine.

Stepping back from solid shores into the oceans of water or other solvents, I propose the following constraint for all life: it is modular, and whatever the modules are, and whatever the solvent is, the modules should be thermodynamically disfavored from reacting indiscriminately with one another in that solvent. A system that has the potential to transition into a living system then needs mechanisms that make selected reactions kinetically accessible and thermodynamically favorable.

**Rebecca Rapf:** I agree with Paul's excellent point about the importance of considering kinetics when considering prebiotic chemistry or origins of life in any environment. Molecules need to be mobile enough to encounter each other for the relevant reactions to occur within the reasonable timescale set by the environment. This brings us back to the limitations posed by experimental timescales (and perhaps our imaginations) discussed above, but it is interesting to consider how the diffusion of chemical species would occur in a solid-state life scenario and what constraints or opportunities this would provide (i.e. perhaps set a natural timescale for life).

**Martin Rahm**: Solid-state diffusion and exchange can occur. For example, we have shown that small hydrocarbons like methane and ethane can diffuse into the solid-state lattice of hydrogen cyanide, a molecule central for the abiotic synthesis of many of (known) life's building blocks.[83]

**Richard Remsing**: Solid-state diffusion can occur through a variety of mechanisms and is extensively investigated in fields like solid-state and polymer electrolytes.[84] One such interesting mechanism of solid-state diffusion involves the rotational motion of one molecule or molecular ion coupling to and facilitating the translational diffusion of another molecule or ion.[85, 86] Though the precise nomenclature is debated (paddle-wheel, soft-cradle, revolving door),[85, 87, 88] the key feature of this diffusion mechanism is the coupling of rotational motion of one species to the translational motion of another.[89, 90] These plastic crystal phases, in which molecules are rotationally disordered, exist over a range of temperatures (depending on the specific material) and could exist in many planetary environments.[69, 91, 92]

**Valentina Erastova:**
Cairns-Smith's proposal that clay crystals acted as the first genes[55] still shapes the way we ask what minerals contribute to early chemistry, as does the related question of whether clays, layered double hydroxides (LDHs), or related layered minerals could template sequence-specific polymers.[59] Both framings may be too narrow.

Rather than acting only as self-replicating sequence-information carriers (AKA "first genes"), *minerals may act as transcribers of environmental information*: the geochemical composition, pH, redox state, and ionic environment of a surface are written into the molecules that adsorb to it and form within it.[93] In this view, mineral interfaces could encode environmental variation in oligomer composition and sequence, allowing the resulting polymers to retain a molecular record of their formation conditions.

This reframing also opens a largely unexplored role for LDHs as proto-protocells. Their layered, charged interlayer spaces provide confinement and concentration, mediate selective adsorption, and can catalyze condensation reactions.[94] However, polymer formation is only half of the problem. The open question is the controlled release: how does a surface-bound oligomer detach in a form that retains the selectivity and potential functionality? This transition from surface-bound to free polymer may be as important as polymer formation itself.

Detection then becomes difficult in a very practical way. Both experimental analogues and extraterrestrial samples would require methods sensitive to low concentrations of short, surface-associated oligomers – a nontrivial analytical challenge that may place some experiments close to or at detection limits. In short, the most interesting chemistry may be the least abundant.

**What kind of life can you imagine at the low temperatures of Titan and similarly cold planetary bodies?**

**Richard Remsing:** Prevailing wisdom suggests low temperature planetary bodies are chemically inert, because the available thermal energy too small to push a chemical system over a reaction barrier and break/form bonds. This classical picture ignores quantum mechanical effects, and quantum reaction pathways can become the dominant reaction mechanism at low temperatures through quantum tunneling.[81, 95] In quantum tunneling, the system does not proceed from reactant to product by moving over a free energy barrier. Instead, the quantum mechanical system can move through the barrier; like a car driving through a tunnel to get from one valley to another instead of driving over the mountain separating them. The probability to tunnel from reactant to product is related to the width of the reaction barrier, and an energetic barrier no longer needs to be crossed for the reaction to occur. At low temperatures, when the available thermal energy is much lower than reaction barriers, quantum tunneling should become the dominant pathway for reactions to occur.[34, 96]

Terrestrial biochemistry seems to leverage quantum tunneling and similar quantum effects whenever possible, even at temperatures on Earth. For example, proton transfer reactions in enzyme active sites often involve quantum tunneling of the proton from reactant to product state,[97] tunneling in proton and electron transfer reactions play important roles in photosynthesis,[98, 99] and it is increasingly apparent that many properties of water, the solvent for terrestrial biochemistry, are significantly impacted by nuclear quantum effects.[100, 101] It is natural to assume that life at much lower temperatures would leverage quantum mechanical effects like quantum tunneling much more than life on Earth, because (1) quantum pathways become the natural way to do chemistry at low temperatures and (2) low temperature environments are more likely to stabilize quantum systems, enabling exploitation of additional quantum effects like coherence to enhance energy transport,[102, 103] sense the surrounding environment,[104] and even enable rapid communication within and between organisms. Life at low temperatures should be quantum.

**Thanja Lamberts:** Hydrogen transfer reactions in life as we know it at Earth's room temperature is already extremely important. One of the reasons that heavy water is poisonous at larger quantities is in fact because it prohibits certain tunneling-mediated reactions. But this also points to where tunneling can be expected to have the highest contributions, namely where the mass involved is low, e.g., electrons and hydrogen atoms. Heavy-atom tunneling is known to play a role too, as long as the larger mass is compensated for by the barrier height and width. A simple way to go about building up knowledge on this may be to structurally report crossover temperatures for the reactions studied, so it is clear whether a tunneling importance cutoff is expected at 20 or 2 K.

In terms of other quantum effects at play at low temperatures, I am not sure which would apply easily here, because of the purity of the materials that have been studied so far. Superfluidity, however, of Helium and para-H2 might make for different kinds of solvents than those that we are used to thinking of. At least for Helium there is a large body of work available experimentally as well.

One way that I tend to think about chemistry at low temperatures is that there needs to be a driving force, say exothermicity. Taking inspiration from plants needing sun light to survive, can we imagine a life form that might require electron cascades brought about by cosmic rays as a chemical driving force? Such radiation might be harmful and dissociative to certain kinds of chemicals available, but that may also offer the potential to act as a way to be selective.

**Dirk Schulze-Makuch:** There are remarkable possibilities for life in a Titan environment. For example, metabolism in an Earth-like environment is based on redox-reactions. These would likely not provide enough energy on Titan, but radical reactions would. They are rarely used in Earth-like life, because they are too energetic and destroy covalent bonds, but under very cold conditions, they could provide sufficient energy. Another example is that there is no rule in biology that life can only be 100 years or perhaps 1000 years old. In principle, it could become a million years old, particularly in a cold environment with little radiation in which not much repair is needed. The downside is that evolution would be very slow and life itself would likely be slow in a cold environment as well.

**Jonathan Lunine:** Leaving aside the important question of the role of quantum tunneling, I imagine "slow life", as Dirk said. It exists on the shoreline of the lakes and possibly seas, relying on surfaces to concentrate the reagents it needs. Perhaps there are cellular membranes (even azotosomes or some equivalent whose formation is enabled by surface chemistry). The shoreline is ideal because it allows access to energy-storing polymers (chemical energy from polymerization of $C_2H_2$, solar energy from polyimines[33]), while the liquid ethane-methane provides the mobility for moving nutrients in and waste out. Maybe the "life" is colonies of complex, interacting molecules that enclose themselves in an azotosome every so often to venture to the bottom of the ethane-methane sea to scavenge organic compounds not available on the shoreline or gather more $C_2H_2$ if the shoreline is depleted. They will need to seek refuge from the drying up of the seas on timescales of 40,000-100,000 years if the precession of Titan's pole moves liquids into the southern hemisphere periodically, as speculated in the literature.

**Ishaan Madan:** On Earth, psychrophiles survive low temperatures through biochemical strategies that preserve reaction rates in water. Cold-active enzymes trade structural stability for catalytic flexibility, with reduced weak interactions and active sites optimized for low temperature catalysis (0 ºC).[105] For Titan, however, liquid water is not a widely available solvent; on the current surface, it is effectively a rock. Therefore, psychrophile extrapolation may not hold at all.

If life dwells in the hydrocarbon lakes, it would likely require a fundamentally non-polar biochemistry. Perhaps energy transduction might occur via electron tunneling or redox chemistry of nitrogen-carbon chains. Information molecules could be small, polar-inert polymers stabilized by van der Waals instead of hydrogen bonds (as discussed during the workshop). Metabolic timescales could be millennia ("slow life" as proposed by Dirk).[106]

I also wonder whether the most interesting Titan environments may not be the bulk liquid lakes themselves, but their interfaces.[15] With water-ice as Titan's bedrock, the water-ice/hydrocarbon-liquid boundary could play a role analogous to rock–water interfaces on Earth. On Earth, mineral surfaces concentrate reactants, catalyze reactions, and create chemical gradients. On Titan, water-ice or other cryominerals may serve a similar role.[69]

Another provocative possibility is that Titan life, if present, may appear as slow-growing colonies, films, crusts, or patterned chemical communities attached to water-ice rocks along lakebeds, shorelines, or transient wetting zones.

**Martin Rahm:** If one, for the sake of argument, assumes that life is present near the surface of Titan, then I would argue it is most likely solid-state life. The reasons are several. The only liquids present are lakes and seas of methane, ethane, and nitrogen. We know what happens when something is dropped into liquid nitrogen: it falls out of solution. The solubility of anything much larger than the solvent molecules themselves is pitifully low, so the analogy with water is quite stretched.

I agree with Jonathan that life on Titan might prefer the shores of the lakes and seas, but I would bet that this life would not be liquid-based. There ought to be little need for such life to manufacture chemical containers, membranes, for there is little risk of dilution of life's machinery. Crystallization is a great way of concentrating chemistry.

Such life might rely on the tides not only to gain low-molecular-weight energetic feedstock ($H_2$, ethane, HCN, acetylene) but also to move itself around. Movement might well be one of the greatest concerns for solid-state life, for it is, by definition, locked in place. The reasons to occasionally move could be several: to find greener pastures, or to disperse following reproduction. This is why I speculated, in my answer to an earlier question, that such life would synthesize explosives. Locomotion through detonation could be a dramatic but effective way of moving from A to an undetermined B on the surface of Titan.

Another, related chemical route that might allow for temporary locomotion is self-melting: a more controlled release of large amounts of stored heat that would permit brief mechanical motion. Yet another way of propelling solid-state life would be through the so-called thermosalient effect, whereby abrupt, anisotropic changes in a crystal lattice cause it to jump. Such phase changes are typically triggered by heat but can also be brought about by light (the photosalient effect). If there are jumping crystals on Titan, they might be made by life.

Solid-state life could be an agglomeration of crystalline and amorphous matter, each serving distinct roles (just as biological cells partition functions across structurally distinct compartments). Crystalline regions would be well suited for information processing, either through controlled defect patterns or by hosting guest molecules within the lattice. Amorphous regions, with a greater structural flexibility, could facilitate chemical reactions and the harvesting of feedstock. Porosity would likely be a feature of both: in crystalline domains, a host lattice accommodating different guest molecules could function as a form of memory; in amorphous regions, porosity would enable diffusion of small feedstock and waste products. A lower overall density would also aid locomotion.

I push this line of thought because such life would be hard to spot. Something even harder to detect would be quantum life: a solid-state life-form that has evolved into something akin to a running program, where complexity resides not in chemical diversity but in electronic structures evolving on picosecond timescales, or in infrared lattice vibrations, far beyond what might be apparent from the simplicity of its chemical substrate. Such an organism might look like nothing more than an imperfect crystal or an organic lump.

**Nita Sahai:** One of the characteristics of life as we know it is that all/most critical reactions are enzymatically-controlled with many steps that have smaller activation energy barriers (~ 20-80 kJ/mol or 5-19 kcal/mol) compared to non-enzymatic or uncatalyzed reactions (~60-100s kJ/mol). Perhaps this is also a key feature of life as we don't know it. Could one then ask how big or small an energy barrier needs to be in order to be considered as a biochemical reaction?

Based on what we know from physical chemistry of life as we know it, the size of enzymatically-catalyzed reaction energy steps likely depends on the amount of available energy in the environment through the dependence on thermal energy ($k_BT$), where $k_B$ is Boltzmann's constant and T is temperature. A comparison of the average kinetic energy of biochemical reactions in psychrophiles versus mesophiles and thermophiles reveals the expected trait that the activation energy of enzymatically-catalyzed reactions is lower than at warmer temperatures[105, 107, 108]. The cold temperatures, however, would also bring additional challenges, such as limiting the flexibility of the enzymes, freezing the intracellular fluid and extracellular solvents, etc. for which the organisms would need adaptations. While the mechanistic pathways for a given enzymatic pathway may be the same, the psychrophilic enzymes are different in both kinetic and thermodynamic features. Specifically, compared to mesophilic and thermophilic counterparts, cold-adapted enzymes have high specific activity at low temperatures, low thermal stability which makes them highly flexible thus preventing slow structural dynamics. Such enzymes also have low activation enthalpy, so they are less sensitive to variations in temperature[107]. Concomitantly, psychrophilic enzymes have a more negative value for the entropy of activation, suggesting more structural disorder in the ground state[107]. The kinetics of folding is ensured by a greater expression of trigger-factors and peptidyl-prolyl *cis/trans* isomerases[107].

For the kind of potential low-temperature life as we don't know it discussed here, some similar principles of numerous catalyzed steps with small energy barriers and thermal-adaptation of inorganic or organic catalyst behaviors/functions may be valid based on their physicochemical logic.

**Rob Hodyss:** I want to emphasize the point touched on by others that Titan surface life (if it exists) could be based on non-covalent chemistry far more than terrestrial life is. At Titan's low temperatures, non-covalent interactions (Van der Waals, CH-pi, pi-pi interactions) become more important. Life may rely on the breaking and making of these kinds of bonds to drive its machinery.

**Rebecca Rapf:** I agree that this is a particularly important point. As mentioned by others, the environment will likely set the useful type of bonding (covalent bonds, van der Waals, etc.) that could potentially be used by life in that environment. Life requires a kind of "sweet spot" between stability and reactivity. Perhaps, by looking to match the energy scale of the ambient conditions, the energy scale can be used as a broader constraint for the possible chemistries that could make up "life as we don't know it" in a given environment. For example, as mentioned by Nita above, it is important to consider the activation barriers for biochemical-like reactions relative to the energy available in the environment. On Earth, the available energy means that covalent bonds, like amide and glycosidic bonds, are relatively long-lived, but not so stable that all chemistry gets "stuck" and can't turn over. Therefore, on Titan or other cold environments, the relevant ambient energy would make it so that non-covalent interactions are our "biochemical" bonds of interest, as there is not enough energy available to break covalent bonds on a reasonable timescale.

**Martin Rahm:** I agree! Non-covalent interactions will be important - maybe even dominate. Imagining dynamic chemistry at very low temperatures challenges our conventional thinking of what is relevant chemistry for life. Figure 3 illustrates how the activation energy of a reaction (or the strength of a chemical bond) translates into the timescale of reaction. The comparison between Earth ambient conditions and Titan is stark: a conventional organic reaction that proceeds over a few minutes at (our) room temperature (298 K) would take longer than the age of the universe at Titan's surface. On the other hand, weak hydrogen bonding interactions can form and break on minute timescales at 94 K. Such interactions (van der Waals less so) are - like covalent bonds - directional, can be structure-determining, and can additionally act catalytically. The chemistry of life, if it exists near the surface of Titan or similarly cold worlds, need not be "slow," provided it is held together by the right kind of interaction.

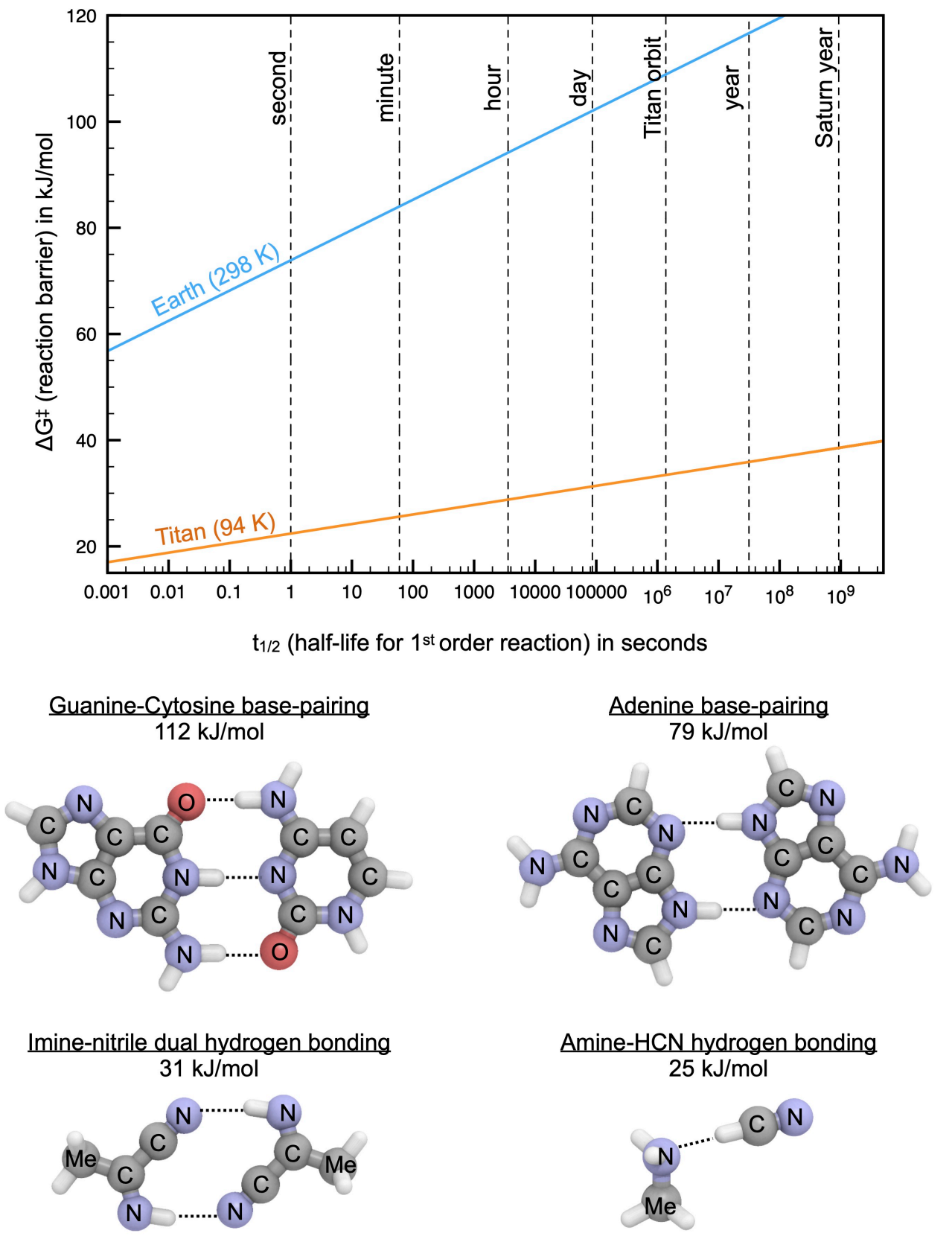

***Figure 3.*** *Chemical life would need to rely on weak interactions to sustain dynamic chemistry at reasonable time scales at extremely low temperatures. Top: Estimates of reaction half-times as function of Gibbs energy of activation are derived from the Eyring equation assuming first-order reaction kinetics. Bottom: Examples of weak interaction motifs based on hydrogen bonding. Binding energies are calculated at the DLPNO-CCSD(T)/aug-cc-pVTZ//wB97X-D3/Def2-SVPD level of theory and are corrected for zero-point energy vibrations. Adapted from Figures 1a and 3 in "The Astrobiology of Titan" from Planetary Astrobiology, edited by Victoria Meadows, Giada Arney, Britney Schmidt, and David J. Des Marais © 2020 The Arizona Board of Regents. Reprinted by permission of the University of Arizona Press.*

**What kind of life can you imagine in the high temperature sulfur/silica environment of Io?**

**Jonathan Lunine:** There is a thorough investigation of alternative solvents for life by Bains et al**,** and they do treat liquid sulfur.[71] However, they point out that the essential properties of sulfur needed to evaluate its suitability (solvent efficacy, reactivity) are not well known, and further, its properties vary widely with temperature. As a consequence, on a body such as Io where temperatures of liquid sulfur are dependent on the primary volcanic fluid, silicates, there may be a wide range of properties over short distances or timescales. It is therefore difficult to know what conditions to assume in order to model the possible sulfur chemistry.

**Martin Rahm:** This might amuse some of you, but if life exists on Io, I will again argue for solid-state life, but of a fundamentally different character than what I imagined for Titan.

On Titan, the case for solid-state life rests on the insolubility of complex organic chemistry in cryogenic solvents. On Io, the situation is very different: the environment appears too energetic, too destructive, for anything delicate to survive. Molten sulfur and silicate lava exist, but these are transient, violent, and extraordinarily hot. Any molecular complexity dissolved in such liquids would decompose rapidly. In other words, the greatest challenge for life on Io is not energy, but persistence. Rock might be the only medium robust enough to persist.

What makes Io chemically interesting, despite its hostility, is sulfur, which Jonathan brought up. Sulfur is one of the most versatile elements in the periodic table in terms of its redox chemistry, accessing oxidation states from −2 to +6, and it forms a large variety of allotropes – rings and polymeric structures, which interconvert as a function of temperature. Ishaan and I have discussed elsewhere in the context of Venus's cloud decks how sulfur (and phosphorus) retain the ability to form polymeric structures even when fully oxidized.[7] This is unlike carbon, which becomes $CO_2$.

I would therefore speculate that solid-state life on Io, if it exists, might inhabit interfaces where both sulfur's chemical versatility and the structural robustness of silicate minerals are available. Sulfur's polymorphic transitions, its ability to form extended structures in multiple oxidation states, and redox gradients presumably present on Io's surface might provide the energetic basis for persistent solid-state chemistry at such interfaces.

Such life would face the same locomotion problem as on Titan, but with a different solution available: on a world this volcanically active, life might not need to move at all. The environment moves for it. Lava resurfacing, volcanic plumes depositing material hundreds of kilometers from its source, and tidal flexing of the crust could serve as mechanisms for dispersal and feedstock delivery. Life that lets itself be buried, redistributed, and re-exposed by geological cycling would be practicing a form of passive dispersal not unlike spore distribution on Earth, but mediated by volcanism rather than wind. Lava tubes are thought to be abundant on Io, and these could provide relatively stable environments shielded from Jupiter's radiation, where temperatures might permit complex chemistry embedded in or on mineral surfaces.

Again, I am worried about detection. If there is life on Io, we would likely not recognize it. It might look like a discolored mineral vein in a lava tube wall, or an unusual crystalline deposit near a volcanic vent.

**Richard Remsing:** Many of the properties of water that we think are special and play an important role in the origin of life are not unique. For example, the tetrahedral arrangement of molecules in water's hydrogen bond network is responsible for ice floating on liquid water and therefore preventing bodies of water from completely freezing.[109, 110] At the molecular scale, the tetrahedral arrangement of water molecules creates a relatively open liquid structure, into which small molecules can be accommodated relatively easily while large molecules break water's tetrahedral network and induce the formation of an interface around the solute molecule. This behavior gives rise to a solute size-dependence of hydrophobic effects that creates driving forces for proteins to fold.[111-113]

Tetrahedral coordination structures are not limited to water but are prevalent in a variety of atomic and molecular liquids, including liquid Si, Ge, $BeF_2$, and $SiO_2$.[114] Due to this tetrahedral structure, these tetrahedral liquids can exhibit many of the same anomalous and often beneficial properties as water.[115-117] As in water, their solid phases often float on the liquid. Solvophobic effects (the generalization of hydrophobic effects to other solvents) exhibit a solute size-dependence that can drive self assembly.[111, 113] Anomalous thermodynamic properties, like a maximum in the density as a function of temperature, may originate from the existence of a phase transition between two liquid phases (low and high density phases).[110, 118-120]

Of course, important differences exist between the various types of liquids. In water, the tetrahedral structure originates from hydrogen bonding driven by an electrostatic attraction between the hydrogen atom of one molecule and the oxygen of another. In contrast, metallic interactions in liquid silicon coexist with transient covalent bonds between Si atoms, such that the presence of covalency creates a tetrahedral network that reorganizes on timescales similar to the hydrogen bond network of water, albeit at the higher temperatures needed to maintain liquid silicon.[121, 122] While liquid Si may be too reactive for typical prebiotic chemistry, its metallic/semimetallic nature would be useful for rapid charge transport.

Liquid silica also exhibits a tetrahedral network, ultimately due to the nature of Si-O covalent bonds. Silica consists of tetrahedral $[SiO_4]^{4-}$ units, in which each oxygen is shared between two Si atoms (on average, defects can exist in the liquid) to create a networked liquid. Silica's open network structure results in many of the same anomalies as liquid water, motivating the question, *could there be alternative prebiotic chemistry or biochemistry in molten rock?* Maybe the most obvious barrier to complex chemistry in molten rock is the high temperatures, at which the resulting thermal energy is much larger than typical ionic, van der Waals, and hydrogen bonding interactions. As a result, maintaining complex chemical structures would need stronger interactions that can be stable at high temperatures, e.g. covalent bonds.

In general, the relative importance of different types of interatomic and intermolecular interactions in alternative biochemistries is set by the thermal energy of the environment, given by the product of Boltzmann's constant and the temperature, $k_B T$. On Earth, when $T$~300 K, the thermal energy is approximately 2.49 kJ/mol. On cold planetary bodies like Titan, the available thermal energy is much lower, approximately 0.78 kJ/mol. In contrast, active lava flows in environments like Io create an excess of thermal energy, $k_B T$~12.5 kJ/mol for $T$~1500 K. Therefore, different types of bonding may play different roles in different environments. On Earth, covalent bonds impart stability and are rarely disrupted by thermal fluctuations, while relatively weaker hydrogen bonds lead to structuring that breaks and reforms on molecular timescales. In cold environments, like Titan, the strength of hydrogen bonds relative to $k_B T$ is similar to the relative covalent bond strength at Earth-like temperatures. This suggests that hydrogen bonds could play the role of covalent bonds in alternative Titan biochemistry, imparting stability on long timescales and even requiring catalysis to break such bonds. At high temperatures typical lava flows, typical covalent bonds would readily break and reform like hydrogen bonds on Earth. Only very strong covalent bonds would have long lifetimes and form stable structures, such that any biochemistry would by necessity be much more dynamic than terran biochemistry.

Sulfur and other liquid chalcogens are also two-state liquids that undergo liquid-liquid phase transitions, and therefore also can exhibit anomalous thermodynamic properties.[123] However, the structure of liquid sulfur is quite different that the tetrahedral liquids discussed above (e.g. water). The liquid phase closest to the melting point consists largely of eight-member rings. At high temperatures, thermal fluctuations open the rings and lead to living radical chain polymerization, creating a polymeric liquid sulfur phase.[124, 125] However, liquid sulfur is apolar (low dielectric constant of 3.5) and solubility of functional polar and ionic macromolecules would present a challenge.

**Thanja Lamberts:** The discussion here makes me wonder whether we could construct – just for the sake of training our imagination – a system, say computationally, that includes such a tetrahedral structure and try out how other chemical compounds behave in that environment. This may give a way to discover how materials and chemicals interact with each other in a way that hasn't been explored before. I wonder whether sufficient force fields are available that might be patched together to allow for a cheap and computational screening of different combinations. I imagine it may be possible to test which kinds of molecules diffuse in the tetrahedral structures mentioned above by Rick and when the material becomes amorphous, and what that might teach us and how it expands our imagination.

**What could be special about life at extremely high pressure?**

**Valentina Erastova:** High pressure may alter the conformational landscape of surface-adsorbed molecules, with consequences for confinement geometry, adsorption selectivity and molecular mobility at mineral interfaces. The question is not merely whether reactions occur under pressure, but whether the selectivity that surface chemistry is assumed to provide is preserved. If high pressure suppresses a required mode of diffusion, assembly, or conformational change, then a living system would need either a pressure-tolerant route to that function or a way to locally alter the pressure.

**Jonathan Lunine:** Take the possibility of a liquid water-ammonia ocean or slushy ice at 10 kbar[126], and consider a putative hydrogen bond-driven chemistry. At such high pressures, the PV term in the free energy definitions would dominate much of the thermodynamics.

**Richard Remsing**: Sufficiently high pressures would disrupt driving forces for Earth-life, especially hydrophobic effects. The solvation of small apolar molecules (volume less than approximately 1 $nm^3$) scales with the solute's volume.[111-113] Larger solutes necessitate the formation of an interface between water and the apolar solute, and the free energy cost of forming this interface (equal to the product of the surface tension and the surface area of the solute) dominates the hydration free energy when the pressure is relatively low (as it is on Earth), such that the PV contribution to the solvation of large apolar solutes is small. This produces solvation free energies that scale with solute volume for small solutes but with solute surface area for large solutes. This change in scaling creates a driving force for many small apolar solutes to assemble into a single large structure, and we call this force hydrophobic effects.[112] At high pressures, the PV term in the solvation free energy will dominate, the solute size-dependent change in scaling no longer occurs – free energies scale with volume for all solute sizes – and there is no hydrophobic driving force for assembly.[127, 128] For example, pressure-induced protein denaturation is well-known[129] and dates back to Bridgman's 1914 paper[130] demonstrating that applying pressure to an egg white makes it look like a hard-boiled egg, ultimately due to pressure-induced denaturation. Life at extremely high pressures would need to leverage alternative driving forces for self-assembly or strategies for locally lowering the pressure to take advantage of hydrophobic effects.

**Thanja Lamberts:** It is useful to think about extremophiles as examples for life at high pressures in Earth environments. Consider tardigrades; at what pressures would they go to "sleep" as pressure is increased?

**Martin Rahm:** One person's high is certainly another bug's low...pressure. Certain organisms have been recorded surviving a couple of GPa of static pressure (1 GPa ≈ 10 000 atm)[131]. Some are even known to survive shock conditions, so for very short timescales, up to 78 GPa[132]. However, at sufficiently high pressures everything becomes solid. And by now you should know where I am going with this. I would rank high-pressure solid environments as the least likely to sustain life though, because they cannot support moving feedstock in and waste out. Diffusion is limited, and porosity is crushed out of existence. I really did not come into this discussion intending to claim that the universe may be full of living rocks. But what if it is?

**How do we transition from prebiotic chemistry to astrobiology?**

**Jonathan Lunine:** Perhaps the question should be rephrased as "How does astrobiological research transition from prebiotic chemistry to chemical modeling of the origin of life". I think we heard some useful connections in the work of Thanja Lamberts on interstellar chemistry. There is a lot of observational evidence from ALMA, and other facilities that constrains the models of interstellar chemistry, along with extensive laboratory simulations. Perhaps the work on interstellar chemistry could serve as a kind of 'best practices' guide for some of the work on chemistry across the protobiological transition to life…whatever that transition might be. I would love to hear from others on this issue!

**Ishaan Madan:** I'm inclined to say that synthetic biology may be one line of transition between prebiotic chemistry and astrobiology. I like to think that astrobiology begins when matter, emerging from prebiotic chemistry, acquires some capacity to maintain itself: preservation, reparation, adaptation, or continuity of organization. But at what level does maintenance emerge? Is it a property of a molecule, a network, a compartment, an environment, or the coupling between all of them? Synthetic biology provides a bridge for testing different aspects of this transition. For instance, recent work has shown that a small RNA motif can synthesize both itself and its complementary strand, approaching two key reactions needed for self-replication.[133] This does not solve the origin of life, but it helps make the boundary experimentally tractable.

Perhaps the transition is not a mere line. Maybe it is a ladder of increasing organization instead: from chemical possibility, to persistence, to information storage, to feedback, to evolution, and eventually to systems that maintain themselves against environmental change[134, 135].

**Nita Sahai:** This question, What is Life, is at the heart of all Astrobiology and OoL research regardless of whether it is Life as We Know it/Don't Know It, and regardless of whether that is on Earth or on other planetary bodies. It is precisely this question that I was trying to discuss during my presentation and discussions all week. If our larger Astrobiology community (and, we, as this workshop group) truly want to answer this question, then we have to face up to philosophical issues of ontology as opposed to the ontics approach, that we have taken as a scientific research field for the last many decades.

It is my view that the question of finding a "Biosignature," whether "Agnostic" (sic) or not, is equally fraught, although some earlier suggestions[136] and recent AI and GC-MS-pyrolysis technological advances seem promising[137, 138]. Again, I would submit that this is because the basic ontology of what is non-living/living has not been completely understood.

The thinking in our community seems to have evolved towards a "fuzzy boundary" between living and non-living. Well, if that is the case, then short of demonstrating abiogenesis of a whole cell with most/many of the features we find in life as we know it, or finding microfossils with chemical/isotope analyses to go along with the microscope images, such as Bill Scopf's, Pilbara Craton, Australia, we won't be able to address important questions. Some of these questions include: How do we transition from prebiotic chemistry to Astrobiology?; What constitutes a Biosignature?; What properties that are presumed necessary for Earth-life may not apply for life as we do not know it? In short, given the speculative nature of Astrobiology, what qualifies as a scientifically good and valid question) with much certitude for the foreseeable future?

**Valentina Erastova:** The observation that biological evolution arrives at equivalent functional solutions via chemically distinct routes (= convergent evolution at the molecular level) provides indirect but important reassurance for prebiotic chemistry research. If life's functional requirements can be met by more than one chemical implementation, then the field is not searching for a unique needle in a haystack but mapping a broader region of accessible chemical space.
This perspective also has practical consequences for detectability: if multiple routes converge on similar functional outputs, then even if we cannot observe the specific pathway taken, we may be able to recognise the functional signature of the outcome. For mineral-mediated chemistry, this reframes geochemical diversity on early Earth – differences in compositions, temperatures, pH, redox state, and cycling -- as a mechanism for sampling multiple convergent routes in parallel. Geochemical diversity is not noise.

**Thanja Lamberts:** Coupling back to the way that astrochemistry has learned to deal with its interdisciplinarity, I think there are some useful ideas to depart from. "We" tend to think along three axes, although I'd advocate for four, being astronomical observations, laboratory experiments, (*ab* initio) theory, and astrochemical modeling. Ideally people working in the field are trained in two out of those, which makes for scientists who develop skills in more than one discipline from the get-go, and this allows them to keep functioning as bridges across the disciplines along their career path. Astrobiology to me appears to have to deal with more than three axes, but even in that higher dimensionality, I suppose it is possible nonetheless to find common denominators and categorize them in one way or another, where perhaps philosophy or ontology should then be an integral part of it. Designing projects in a way that covers more than one axis might help to move forward in a somewhat coherent way, by having many interconnected people and research ideas.

## CONCLUSIONS

Ten years ago, the question of "weird life" – on Titan, on Venus, in environments with no terrestrial analog – had received far less sustained attention than it deserves. What emerged from five days of discussion in Telluride was a clearer picture of where the frontier of this subject lies. The chemistry of life as we don't know it becomes scientifically tractable only when speculation is constrained by planetary context. Environmental constraints such as the thermal energy available, the solvent or its absence, the timescales of reaction and of environmental change, and the abiotic background against which anything living must be recognized are necessary to guide both theory and experiment.

Astrobiology is less speculative than it has ever been, because hypotheses can now be tested on decadal rather than generational timescales. The 2030s should be the decade of the large moons of the outer solar system: Europa Clipper arrives at Jupiter in 2030,[29] JUICE reaches Ganymede and Callisto in 2031,[30] and Dragonfly is to launch in 2028 and arrive at Titan in 2034.[31] The Perseverance rover has cached samples of early Mars whose return would transform our view of that planet's habitability. Several missions to Venus, including ESA's EnVision orbiter, launching in 2031[139], NASA's DAVINCI, whose descent probe is to reach the atmosphere in 2033[140], NASA's

VERITAS, launching no earlier than 2031[141, 142], and Rocket Lab's privately funded Venus Life Finder probe[143, 144], will deliver the first in situ and orbital measurements of Venus's atmospheric and surface chemistry in four decades. Despite the promise that these missions bring to planetary science, chemistry of life as we don't know it as discussed here, especially the possibility of life at high and low temperatures or in the solid-state, will likely be undetectable. The chemical signatures of life as we don't know it should be considered throughout the mission design process.

To look for life as we don't know it, we first need to conceive of the possibilities. This is a task for coordinated, interdisciplinary, and international teams, and it is one for which astrobiology is uniquely well placed. Interdisciplinary efforts of substantial scale can create the flexibility to pursue open-ended questions of the kind discussed here. Our workshop was a success in this sense, a highly interdisciplinary group of scientists came together to discuss and refine ideas that would not have been produced alone, resulting in new directions and strategies for recognizing the chemistry of life as we don't know it.

## AUTHOR INFORMATION

### Corresponding Author

* MR: martin.rahm@chalmers.se; RCR: rick.remsing@rutgers.edu

### Present Addresses

†If an author's address is different than the one given in the affiliation line, this information may be included here.

### Author Contributions

The manuscript was written through contributions of all authors. All authors have given approval to the final version of the manuscript.

## ACKNOWLEDGMENT

MR acknowledges The Swedish Research Council (grant 2024-05049). RCR was supported by the U.S. National Aeronautics and Space Administration under Grant No. 80NSSC22K1132. IM expresses his gratitude to MR and RCR for an invitation to the workshop. As a graduate student, such an experience was invaluable. IM was supported by his advisor, Dr. Ben K. D. Pearce and the Frederick N. Andrews Fellowship awarded by Purdue University. NS is grateful for financial support from NSF EAR 829695, CIFAR Earth4D Fellowship, CIFAR Catalyst CF-0376 - CP24-054, NASA ICAR 80NSSC21K0595 (as a Consultant), and a generous private donor. RJR acknowledges support from the Welch Foundation (W-2156). Part of this work was carried out at the Jet Propulsion Laboratory, California Institute of Technology, under a contract with the National Aeronautics and Space Administration.